\documentclass[         %
aps,                    %  American Physical Society
jpa,                    %  Journal Physics A
showpacs,               %  Displays PACS after abstract
nofootinbib,            %  Does not treat footnotes as references
showkeys,               %
preprintnumbers,        %
floatfix]               %  Fixes float errors
{revtex4}               %  REVTEX 4 Package used
\tighten
\usepackage{graphicx} 
\begin{document}

\title{The Effect of Sterile States on the Magnetic Moments of Neutrinos}

\author{A.B. Balantekin and N. Vassh}
  
  \address{Physics Department, University of Wisconsin, Madison WI 53706 USA}

\begin{abstract}
We briefly review recent work exploring the effect of light sterile neutrino states on the neutrino magnetic moment as explored by the reactor and 
solar neutrino experiments. 
\end{abstract}

\maketitle

\section{Introduction}

Several anomalous results, some more recent,  of the neutrino experiments can be interpreted as consequences of the mixing of light sterile states with the active ones. The existence of such a light sterile neutrino indeed has many interesting consequences \cite{Abazajian:2012ys}. Sterile states cannot be produced in weak interactions, but  they can in principle have sizable electromagnetic couplings through loop corrections in the effective Lagrangians. For both active and sterile states the electromagnetic coupling is usually expressed as neutrino dipole moments. In previous work \cite{Balantekin:2013sda} we elaborated on the effects of light sterile neutrinos on the effective electron neutrino magnetic moment measured at the reactors and showed that the kinematical effects of the neutrino masses are negligible even for light sterile neutrinos. Here we expand on that discussion by providing more details and also including measurements with other sources of neutrinos such as the Sun.

\section{Mass Eigenstate Neutrino-Electron Scattering}

\begin{figure}[t]
\includegraphics[scale=0.17]{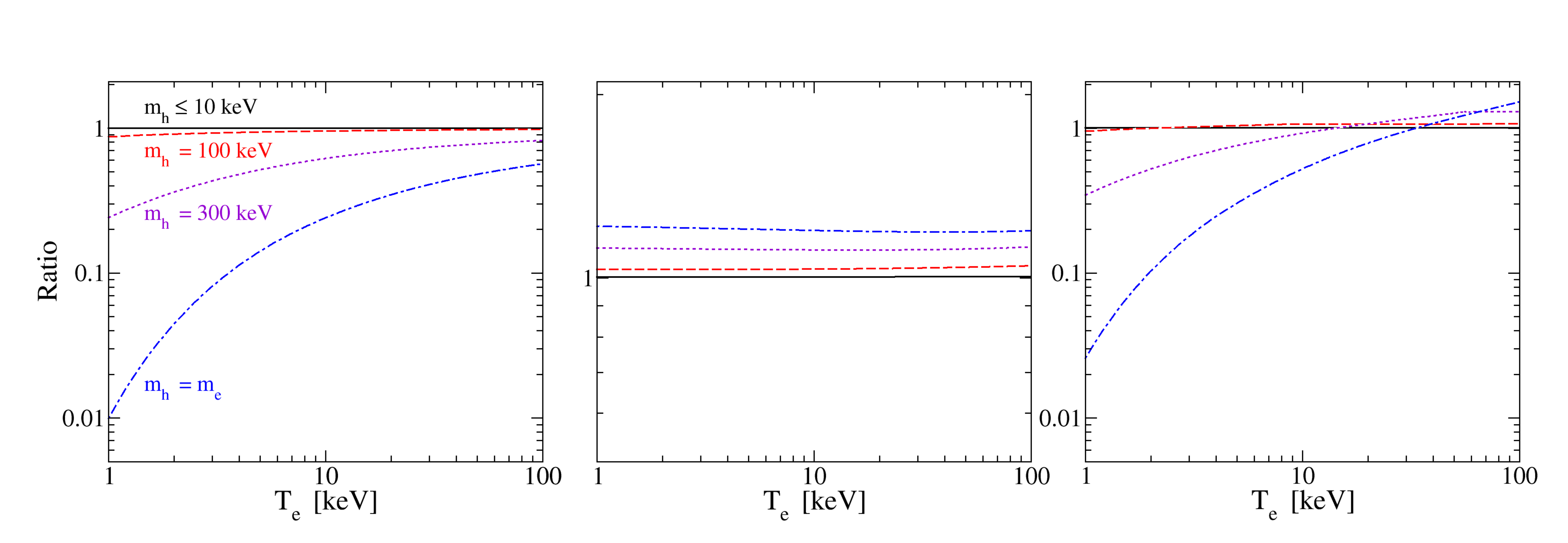}
\caption{(Color online) Left plot corresponds to case $m_{i}<<m_{j}=m_{h}$ with $m_{i}\rightarrow 0$. The center plot is for the case $m_{i}=m_{j}=m_{h}$. The plot on the right is for the case $m_{j}<<m_{i}=m_{h}$ with $m_{j}\rightarrow 0$. In all plots the solid line is for $m_{h}\leq 10$ keV, the dashed line corresponds to $m_{h}=100$ keV, the dotted line represents $m_{h}=300$ keV, and the dashed-dotted line shows the result for $m_{h}=m_{e}$.}
\label{fig:fig01}
\end{figure}

The differential cross section for scattering from neutrino mass eigenstate with mass $m_i$ to the mass eigenstate with mass $m_j$ 
is \cite{Balantekin:2013sda}
\begin{equation}
\label{1}
\frac{d\sigma_{ij}}{dT_{e}} = \frac{e^2 \mu_{ij}^2}{16 \pi m_{e}}\frac{1}{E_{\nu}^{2}-m_{i}^{2}} \left[ \frac{4 m_{e} E_{\nu}^{2}}{T_{e}} - 4 m_{e}E_{\nu} + \Delta\left(\frac{2E_{\nu}}{T_{e}}-1\right) + \Delta^{2}\left(\frac{1}{2 m_{e} T_{e}} - \frac{1}{2 T_{e}^{2}}\right) - \frac{m_{e}}{T_{e}}\left(m_{i}^{2}+ m_{j}^{2}\right) \right]
\end{equation}
where $\mu_{ij}$ is the neutrino magnetic moment connecting the mass eigenstates $i$ and $j$, and $\Delta = m_i^2-m_j^2$. 
In writing Eq. (\ref{1}) it is assumed that electron is initially at rest with  $E_{\nu}$ and $T_e$ being the total energy of the incoming neutrino and the final kinetic energy of the electron, respectively. 
Since reactor experiments searching for the neutrino magnetic moment measure the recoil energy of the electron struck by neutrinos with energies between a few eV to about 10 MeV, the quantity most appropriate to assess the effect of non-zero neutrino mass is the folded differential cross section given by
\begin{equation}
\label{2}
\left<\frac{d\sigma}{dT_{e}}\right> = \int_{E_{\nu}^{min}(T_{e})}^{\infty}\frac{dN_{\nu}}{dE_{\nu}}\frac{d\sigma(E_{\nu})}{dT_{e}}dE_{\nu}
\end{equation}
where we take $dN_{\nu}/dE_{\nu}$ to be the antineutrino spectrum due to fissioning $^{235}U$ given in \cite{Vogel:1988}. Note that the spectra in Ref. \cite{Vogel:1988} are applicable for $E_{\nu} \ge 1.8$  MeV. For energies less than $1.8$ MeV, we use a  power law modification to the spectrum which averages over the peaks mentioned in \cite{Vogel:1988}. Additionally since here we are considering massive neutrinos, the spectrum was taken to be a function of the neutrino kinetic energy rather than the total energy. The kinematically allowed $E_{\nu}^{min}$ for a given $T_{e}$ can be found from the energy-momentum conservation condition to be
\begin{equation}
\label{3}
E_{\nu}^{min} = \frac{1}{2}T_{e}+\frac{m_{j}^{2}-m_{i}^{2}}{4m_{e}} + \frac{1}{4m_{e}T_{e}}\sqrt{(T_{e}^{2}+2m_{e}T_{e})(m_{i}^{4}-2m_{i}^{2}(m_{j}^{2}-2m_{e}T_{e})+(m_{j}^{2}+2m_{e}T_{e})^{2})} .
\end{equation}
We explore the folded differential cross section for three possible scenarios: $m_{j}=m_{h}\gg m_{i}$ with $m_{i}\rightarrow 0$ , $m_{i}=m_{j}=m_{h}$, and $m_{i}=m_{h}\gg m_{j}$ with $m_{j}\rightarrow 0$ where $m_{h}$ is the heaviest of the mass eigenstates. For these three cases we plot the quantity
\begin{equation}
\label{fracratio}
\left[\int_{E_{\nu}^{min}(T_{e})}^{\infty}\frac{dN_{\nu}}{dE_{\nu}}\frac{d\sigma_{ij}}{dT_{e}} \left( m_i \neq 0 \>\>\> {\rm and/or} \>\>\> m_j \neq 0 \right) dE_{\nu}\right] 
\Bigg/
\left[\int_{E_{\nu}^{min}(T_{e})}^{\infty}\frac{dN_{\nu}}{dE_{\nu}}\frac{d\sigma_{ij}}{dT_{e}} \left( m_i = 0 = m_j \right) dE_{\nu} \right]
\end{equation}
i.e., the ratio of the folded differential cross section with non-zero mass to that with zero mass, in Figure \ref{fig:fig01}.

Clearly even for neutrino masses which are quite large ($m_{\nu}\sim 10$ keV) the folded differential cross section does not appreciably change from the massless neutrino limit and we can safely take neutrino masses to be zero in Eq. (\ref{1}). In the limit $m_i \sim 0 \sim m_j$ Eq. (\ref{1}) takes the familiar form:
\begin{equation}
\label{7}
\frac{d\sigma_{ij}}{dT_e} = \frac{\alpha^2 \pi}{m_e^2} \kappa_{ij}^2 \left[ \frac{1}{T_e} - \frac{1}{E_\nu} \right] ,
\end{equation}
where $\kappa_{ij}$ is the magnetic moment in units of Bohr magnetons.

From Figure \ref{fig:fig01} it can been seen that if one is interested in a large mass sterile state in the range ($10$ keV $< m_{\nu} \leq m_{e}$) for the case of a light neutrino transforming into a heavy neutrino (left) or the case of a heavy neutrino transforming into a light neutrino (right) the folded differential cross section is actually reduced by the presence of a heavy neutrino mass state rather than enhancing the value to be more experimentally visible over the weak interaction background. Although the case of a heavy incoming and outgoing neutrino (center) does show an enhancement which gets consistently larger with heavier mass, the result still does not change significantly from the zero mass case.  

It is interesting to note that for the case $m_{i}=m_{h}\gg m_{j}$ with $m_{j}\rightarrow 0$ a divergence in the differential cross section exists for a given $m_{i}$ since the minimum initial kinetic energy of the neutrino goes to zero at 
\begin{equation}
\label{8}
T_{e,c} = \frac{m_{i}^{2}}{2(m_{i}+m_{e})}
\end{equation}
which can be seen from examining Eq. (\ref{3}). However since the spectrum vanishes at this energy, this behavior  is smoothed over in the folded differential cross section and therefore negligible. 

\section{Standard Model Effective Neutrino Magnetic Moment with Three Flavors}

In experiments measuring the neutrino magnetic moment, one needs to take into account oscillations between the source and the detector over the distance $L$, leading to an incoherent sum of the individual cross sections \cite{Grimus:1997aa,Beacom:1999wx}:  
\begin{equation}
\label{8a}
\frac{d\sigma}{dt} = \frac{e^2}{4 \pi} \sum_i \left| \sum_j U_{ej} e^{-iE_jL} \mu_{ji}\right|^2
\left[ \frac{1}{t} + \frac{1}{s-m_e^2} \right].
\end{equation}
where the final neutrino states are summed over since they are not observed in such experiments.  Eq. (\ref{7}) then becomes
 \begin{equation}
\label{9}
\frac{d\sigma}{dT_e} = \frac{\alpha^2 \pi}{m_e^2} \mu_{\rm eff}^2 \left[ \frac{1}{T_e} - \frac{1}{E_\nu} \right] ,
\end{equation}
where $\mu_{\rm eff}$ is the effective neutrino magnetic moment measured at a distance $L$ from the neutrino source and written in units of Bohr magneton. In writing down Eq. (\ref{8a}) matter effects are ignored. If the matter is present, the factors $e^{-iE_jL}$ should be replaced with the appropriate amplitudes calculated using the evolution equations for neutrino propagation in matter. 
If one considers active flavors only, in the case of reactor neutrinos this effective magnetic moment is given by 
\begin{equation}
\label{effmur}
\mu_{\rm eff, reactor}^2 = \sum_{i=1}^3 \left| \sum_{j=1}^3 U_{ej} \mu_{ji}\right|^2, 
\end{equation}
since at short distances the oscillating term can be ignored for active neutrinos, such as the case for the detectors of GEMMA \cite{Beda:2013mta} and TEXONO \cite{Deniz:2009mu} reactor neutrino experiments measuring the neutrino magnetic moment. 
However if there is a fourth (sterile) state then it oscillates and changes the composition of the total flux by the time neutrinos reach the distances where detectors measuring electron recoil are placed \cite{Balantekin:2013sda} yielding 
\begin{equation}
\label{zz1}
\mu_{\rm eff, reactor}^2 = \sum_{k=1}^3 \left| \sum_{i=1}^3 U_{ei} \mu_{ik} \right|^2  +  \left| \sum_{i=1}^3 U_{ei} \mu_{i4}  \right|^2 + \left| U_{e4} \right|^2 \sum_{i=1}^3 \mu_{i4}\mu_{4i} , 
\end{equation}
where we have assumed that neutrinos are Majorana particles.  Clearly such sterile states can be repopulated by the scattering process. 

We are interested in the values for the reactor and solar effective magnetic moments predicted by the Standard Model (minimally extended to include a massive neutrino).  For Dirac neutrinos this prediction is \cite{Fujikawa:1980yx} 
\begin{equation}
\label{10}
\mu_{ij} = - \frac{e G_F}{8\sqrt{2}\pi^2} (m_i+m_j)  \sum_{\ell} U_{\ell i} U^*_{\ell j} f(r_{\ell}) 
\end{equation} 
with
\begin{equation}
f(r_{\ell}) \sim -\frac{3}{2} + \frac{3}{4} r_{\ell} + \cdots, \>\>\> r_{\ell}= \left( \frac{m_{\ell}}{M_W}\right)^2.  
\end{equation}
For Majorana neutrinos only non-diagonal magnetic moments are permitted. In the case the CP-eigenvalues of the two neutrinos are opposite, the neutrino electric dipole moment is zero and the non-diagonal terms of the neutrino magnetic dipole moment are given by Eq. (\ref{10}), multiplied by a factor of 2 \cite{Pal:1981rm}.

\begin{figure}[h]
\includegraphics[scale=0.22]{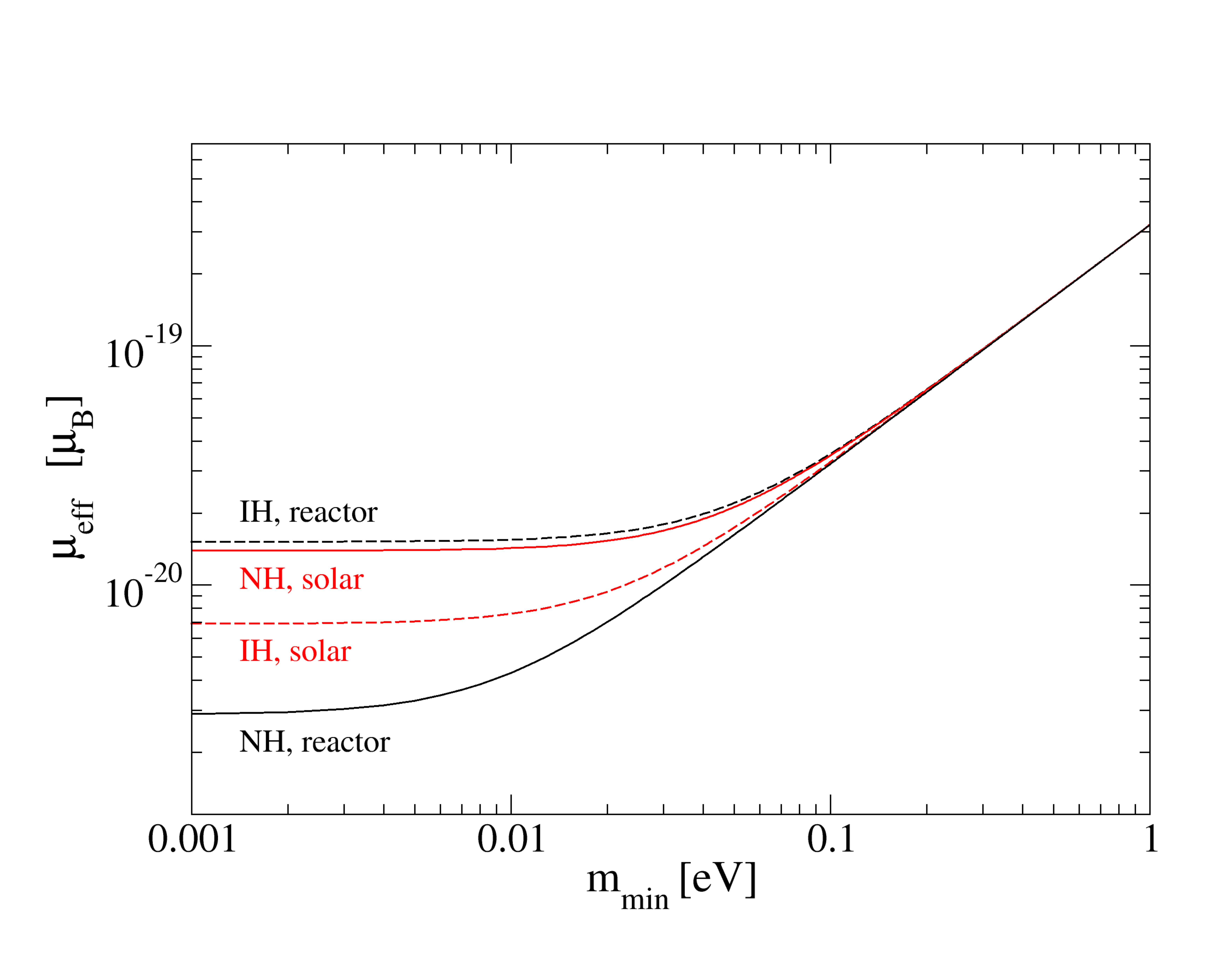}
\caption{(Color online) Standard Model prediction for the effective magnetic moments of Dirac neutrinos for both the solar and reactor case. NH (solid) and IH (dashed) denote normal and inverted hierarchies respectively. The values for the mass splittings and mixing angles were taking from the compilation of the Particle Data Group \cite{Beringer:1900zz}.}
\label{fig:fig1}
\end{figure}

Since the neutrino mass differences and all the mixing angles were recently measured with good accuracy, one can calculate the Standard Model prediction of the effective neutrino magnetic moment as a function of the smallest neutrino mass. For reactor experiments aiming to measure neutrino magnetic moment this prediction was calculated in Ref. \cite{Balantekin:2013sda}. In the case solar neutrinos, Super-Kamiokande collaboration searched for distortions to the energy spectrum of recoil electrons arising from magnetic scattering due to a nonzero neutrino magnetic moment. From the absence of a clear signal they conclude $\mu_{\nu} \le 3.6 \times 10^{-10} \mu_B$ at 90 \% C.L. \cite{Liu:2004ny}. Here we present results for effective neutrino magnetic moment measured using solar neutrinos on Earth. We calculate 
\begin{equation}
\label{effmus}
\mu_{\rm eff, solar}^2 = \sum_i \left| \sum_j \left(A_eU_{ej}+A_{\mu}U_{\mu j}+A_{\tau}U_{\tau j}\right) \mu_{ji}\right|^2, 
\end{equation}
where $A_x$ is the amplitude of the flavor state $\nu_x$ in the solar neutrino flux reaching the terrestrial detectors. The result for the Dirac neutrinos is given in Figure \ref{fig:fig1} as a function of the lowest neutrino mass. 

\begin{figure}[h] 
\includegraphics[scale=0.22]{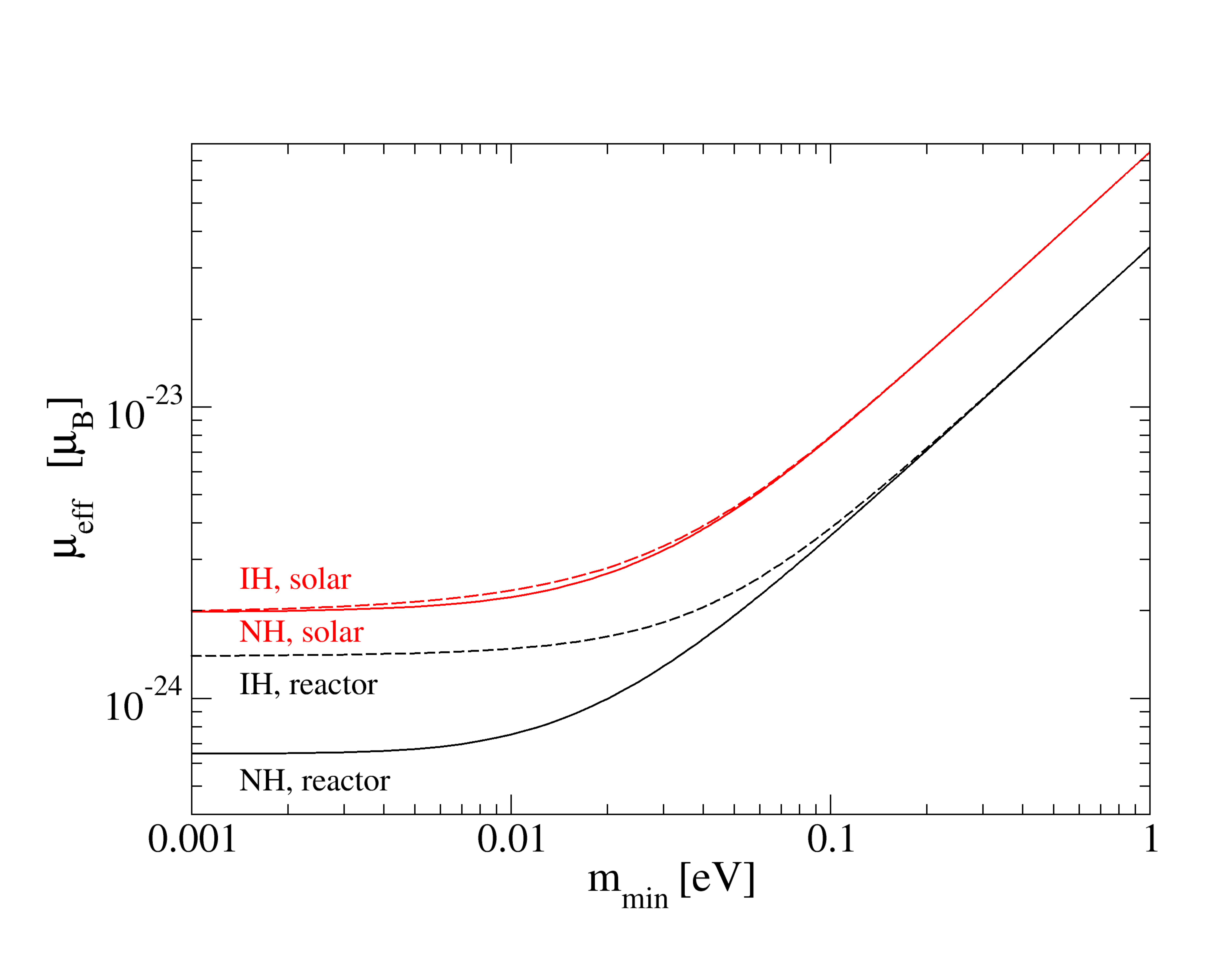}
\caption{(Color online) Same as for Fig.\ref{fig:fig1} but for Majorana neutrinos. Here all Majorana phases are taken to be zero. }
\label{fig:fig2}
\end{figure}

Standard model predictions for the Majorana neutrino magnetic moment as a function of the lowest neutrino mass is given in Figure \ref{fig:fig2}. These predictions are much lower than those for Dirac neutrinos. This is because in the Standard Model the diagonal contribution to the neutrino magnetic moment is dominant whereas the non-diagonal contributions are suppressed by a GIM-like mechanism (cf. Eq. \ref{10}). 

From examining Figure \ref{fig:fig1}  it is clear that the Standard Model prediction for the effective neutrino magnetic of solar and reactor neutrinos are identical for the case of a Dirac neutrino with m$_{min}$ in the range of $0.1$ to $1$ eV. In the lower mass range, a difference between solar and reactor neutrinos cannot be easily discerned from the effect of normal vs. inverted hierarchy. From Figure \ref{fig:fig2} it is evident that for the Majorana case the prediction of the Standard Model for the effective magnetic moment of solar neutrinos is always slightly higher than that for reactor neutrinos, but is still within the same order of magnitude. For both Dirac and Majorana neutrinos the Standard Model prediction is well below the experimental limit.

\section{Conclusions}

The light sterile neutrino explanation of the neutrino anomalies is debatable and clearly needs to be experimentally tested \cite{Djurcic:2013oaa}. 
However, its broadest implications should also be explored.  We presented a brief analysis of the impact of sterile states on the neutrino magnetic moment. The Standard Model prediction for this quantity is too small to be experimentally accessible in the foreseeable future. However, if indeed light sterile states that mix with the active ones are discovered and neutrino magnetic moment experiments measure a value larger than the Standard Model prediction, then unfolding contributions from active and sterile states will provide us another tool to uncover properties of neutrinos. 

%%%%%%%%%%%%%%%%%%%%%%%%%%%%%%%%%%%%%%%%%%%%%%%%
%% BACKMATTER
%%%%%%%%%%%%%%%%%%%%%%%%%%%%%%%%%%%%%%%%%%%%%%%%

This work was supported in part 
by the U.S. National Science Foundation Grant No.  PHY-1205024, in part by the University of Wisconsin Research Committee with funds granted by the Wisconsin Alumni Research Foundation.

%%%%%%%%%%%%%%%%%%%%%%%%%%%%%%%%%%%%%%%%%%%%
%% MAINMATTER
%%%%%%%%%%%%%%%%%%%%%%%%%%%%%%%%%%%%%%%%%%%%

\end{document}